\documentclass{aa}
\usepackage{epsfig}
\begin{document}

   \title{VLBI observations of T Tauri South}

   \author{Kester Smith\inst{1,2,3}
          \and
          Michele Pestalozzi\inst{4}
          \and
          Manuel G\"udel\inst{2}
          \and
          John Conway\inst{4}
          \and
          Arnold O. Benz\inst{1}
          }

   \offprints{K. Smith}

   \institute{Institut f\"ur Astronomie, ETH-Zentrum,
              CH-8092 Z\"{u}rich, Switzerland. 
              \and
              Paul Scherrer Institut, W\"{u}renlingen und Villigen,
              CH-5232 Villigen PSI, Switzerland. 
              \and
              Max Planck Institut f\"ur Radioastronomie,
              Auf dem H\"ugel 69,
              D-53121 Bonn,
              Germany.
              \and
              Onsala Space Observatory, 43992 Onsala, Sweden \\ \\
              kester@mpifr-bonn.mpg.de \\
              michele,jconway @oso.chalmers.se \\ 
              guedel, benz @astro.phys.ethz.ch \\ }

   \date{Received  ; accepted }

   \abstract{We report observations of the T Tauri system at 8.4~GHz
with a VLBI array comprising the VLBA, VLA and Effelsberg 100m
telescopes. We detected a compact source offset approximately 40~mas
from the best infrared position of the T~Tau~Sb component. This source
was unresolved, and constrained to be less than 0.5~mas in size,
corresponding to 0.07~AU or 15~R$_{\odot}$ at a distance of
140~pc. The other system components (T~Tau~Sa, T~Tau~N) were not
detected in the VLBI data.  The separate VLA map contains extended
flux not accounted for by the compact VLBI source, indicating the
presence of extended emission on arcsecond scales.  The compact source
shows rapid variability, which together with circular polarization and
its compact nature indicate that the observed flux arises from a
magnetically-dominated region. Brightness temperatures in the MK range
point to gyrosynchrotron as the emission mechanism for the steady
component.  The rapid variations are accompanied by dramatic changes
in polarization, and we record an at times 100\% polarized component
during outbursts. This strongly suggests a coherent emission process,
most probably an electron cyclotron maser. With this assumption it is
possible to estimate the strength of the local magnetic field to be
1.5-3 kilogauss.  \keywords{Stars: pre-main sequence, Stars: magnetic
fields, Radio continuum: stars}}

   \maketitle
%

\section{Introduction}

Classical T Tauri stars are still in the process of contraction
towards the zero age main sequence, and continue to accrete material
through circumstellar disks. The inner part of the disk is believed to
be truncated by strong magnetic fields anchored on the star, and the
disk material may then be channeled down field lines onto the stellar
surface. The accretion process is thought to be linked to the launch
of highly-collimated jets from the inner disk or magnetosphere.
Evidence for this scenario comes primarily from observations of
infalling material close to the star, and measurements of kilogauss
strength photospheric fields for a few objects (Basri, Marcy \& Valenti, 1992,
Guenther et al. 1999, Johns-Krull et al. 1999a, b). 

In the solar corona, magnetic fields lead to non-thermal radio
emission by various processes caused by accelerated electrons
(see e.g. review by Benz 2002).  Particle acceleration is a basic process
in flares, probably initiated by reconnection of field lines in active
regions. Some of these accelerated particles (those with energies
typically 20-100 keV), can ablate material from the solar surface,
leading to the emission of soft X-rays. In the solar interior, free
magnetic energy is built up by subsurface convection. The fields
envisaged for T Tauri stars probably also arise due to
convection, but are thought to have typically a larger length scale
than those of the Sun, and probably more resemble the large-scale
fields thought to exist in RS CVn systems (see e.g. White 2000 and
references therein). Nevertheless, the solar analogy is instructive in
phenomena such as flares, where the same physical processes lie behind
the observed behaviour. Disks may be important for magnetic energy
release in very young stars since, if the inner disk and the stellar
surface are not rotating in phase, then magnetic fields anchored at
both sites will be wound up and transfer angular momentum between star
and disk (Montmerle et al. 2000). This will also lead to magnetic
reconnection and energy release after a sufficient amount of rotation
(Hayashi, Shibata \& Matsumoto, 1996).  Such processes may also play a
role in driving jets and bipolar outflows.

T~Tauri itself is a highly complex multiple system. The northern
optical source (hereafter T~Tau~N) was found to have an infrared
companion, T~Tau~S, approximately 0.7$''$ to the south (Dyck, Simon \&
Zuckerman, 1982, Ghez et al. 1991).  T~Tau~S has been found also to be
binary, with a separation of approximately 100~mas (Koresko 2000,
K\"ohler, Kasper \& Herbst 2000). Subsequent observations of the two
southern sources (hereafter referred to as Sa for the eastern object
and Sb for the western) revealed relative motion, which could be a
segment of an orbit (Duch\^ene, Ghez \& McCabe 2002). {\em K}-band
spectroscopy by these authors reveals that T~Tau~Sb has spectral type
M, whereas T~Tau~Sa is a continuum source, having an SED probably
peaking at IR wavelengths. The IR observations show that T~Tau~Sa
remains nearly stationary, indicating that it is the most massive
object in the southern system. Johnston et al (2003) observed the
T~Tau system at 2cm with the VLA in `A' configuration between 1983 and
2001. They detected both the northern and southern components. By
assuming that the southern radio component could be identified with
the M star, they were able to determine an orbit for the Sa-Sb pair
and hence estimate the total mass of the southern
system to be 5.3 $M_{\odot}$.

Other radio observations at high resolution have revealed important
polarization and variability features at all observed wavelengths
(e.g. Philips, Lonsdale \& Feigelson, 1993, at 18\,cm using VLBI and
VLA; Skinner \& Brown, 1994, at 2, 3.6 and 6\,cm using the VLA in A
configuration; Ray et al., 1997, at 6\,cm using MERLIN). T~Tau~S
dominates the emission at radio frequencies. Skinner \& Brown found
T~Tau~S to have a falling spectrum and circular polarization,
suggestive of a magnetic origin for the emission. They also observed a
50\% flux increase, accompanied by a reversal in polarization, from
left- to right-handed. Ray et al. (1997) claim to have detected
polarized emission associated with T~Tau~S, with regions of left and
right handed circular polarization on either side of the star offset
from one another by some 0.15$''$ along a northwest-southeast axis.
This was interpreted as emission from a magnetically-threaded outflow,
with oppositely-directed field lines leading to opposite observed
circular polarization. On a slightly larger scale, White (2000)
reported high resolution VLA observations showing extended structure
to the northeast of the southern component.

It has been suggested that T~Tau~N is seen roughly pole on to the line
of sight, based on a comparison of the photometric period, the stellar
radius and the rotational broadening of photospheric lines.  This
suggestion has been used to explain the discrepant extinction between
T~Tau~N and~S, with the northern component viewed through a cavity in
the envelope cleared by the outflow (Momose et al., 1996).  The jet
from the T~Tau~S system, however, appears to lie nearly in the plane
of the sky (van Langevelde et al. 1994), a view reinforced by
modelling of the scattered optical and NIR light surrounding the stars
(Wood et al., 2001). If this is so, the T~Tau components must have
strongly misaligned rotation axes.

In this paper, we present VLBI observations of the T~Tau S 
system.  Our main intention was to probe the inner regions of the
system in order to study the collimating mechanism of the jet at
higher resolution than achieved by Ray et al.  We serendipitously
observed the T~Tau~S system undergo two strong flux increases, accompained by
changes in circular polarization. The extra flux in the second of
these brightenings appears to be 100\% circularly polarized, something
which to our knowledge has never been seen previously in an accreting
pre-main sequence system. The spatial resolution offered by the VLBI
interferometry network should enable us to separate the two components
of T~Tau~S, but the maps reveal only one source. We identify our
detection as being T~Tau~Sb, the M star of the southern binary
system.  We also fail to detect T~Tau~N at high resolution, despite
detecting it clearly with the VLA. In the further discussions, we will refer
to the entire T~Tau~S system when the single components are not explicitly
named.

\section{The observations}

The observations were conducted on December 15th, 1999. All times
referred to are measured from $0^{h}$UT on this date. Ten VLBA
stations were used, together with the Effelsberg 100m telescope and
the phased VLA in `B' configuration. The observing frequency was
8.4GHz (3.6cm, X band). The bright quasars J0431+1731 and J0357+2319,
lying at 3$^{\circ}$ and 7$^{\circ}$ from T~Tau respectively, were
used as phase calibrators.  Since astrometry was not the primary goal
of the experiment, phase calibrators with positions determined from
geodectic observations were not required. The two phase calibrators
used had positions determined from VLBA survey observations and 
were disposed along a line with T~Tau between them.
The phase solutions on both of them should have allowed us to get
precise phase solutions at the target location. Unfortunately the
observing schedule contained too few scans on the secondary calibrator
making the signal to noise ratio on J0357+2319 too bad to get reliable
phase solutions.  J0357+2319 was therefore used to check the quality
of the phase referencing.  Hybrid maps of J0357+2319 were produced
showing an offset from the phase center of approximately
20\,mas. Since T~Tau~S lies between the two calibrators, we can be
sure that the astrometric accuracy at the T~Tau~S position is at least
20\,mas, and probably better in proportion to the relative proximity
of T~Tau~S to the primary phase calibrator. This will be discussed
further in Section~\ref{sourcepositions}.

The T~Tau~S observations were scheduled in 9 blocks consisting of three 
alternations between T~Tau~S (3 minutes) and J0431+1731 (2 minutes), with
one scan on J0357+2319 (2 minutes) at the beginning and the end of each
block. The total observation time on T~Tau~S was about 1.5 hours. The
blocks were separated by gaps of approximately one hour, during which we
observed another two targets which are not discussed in the present paper.

\section{Results}

\subsection{Source positions}
\label{sourcepositions}

\begin{table*}
\caption{\label{positions} Source positions. All positions have been
proper motion corrected to the observation epoch, as described in the
text, but no attempt has been made to correct for orbital motion.  The
uncertainty in the last digit is given in parentheses.  References:
{\em (a)} Hipparcos position proper motion corrected to observation
epoch with Hipparcos proper motion.  {\em (b)} Offset from T~Tau~N given
by Duch\^ene et al. (2002). {\em (c)} Offset from T~Tau~Sa given by
K\"ohler et al. (2000).  {\em (d)} Johnston et al. (2003).  }
\begin{center}
\begin{tabular}{|l|l|l|l|} \hline
Source                & \multicolumn{1}{|c|}{$\alpha$} & \multicolumn{1}{|c|}{$\delta$}        & Ref      \\ \hline
T Tau N (opt)         & 4$^h$  21$^m$  59$^s$.434 (1)  &  19$^{\circ}$  32$'$ 06$''$.429 (14)  & {\em a}  \\ 
T Tau Sa (IR)         & 4$^h$  21$^m$  59$^s$.435 (1)  &  19$^{\circ}$  32$'$ 05$''$.727 (14)  & {\em b} \\
T Tau Sb (IR)         & 4$^h$  21$^m$  59$^s$.427 (1)  &  19$^{\circ}$  32$'$ 05$''$.722 (14)  & {\em b} \\
T Tau Sb (IR)         & 4$^h$  21$^m$  59$^s$.429 (1)  &  19$^{\circ}$  32$'$ 05$''$.704 (14)  & {\em c} \\
T Tau S (2cm)         & 4$^h$  21$^m$  59$^s$.4242 (1) &  19$^{\circ}$  32$'$ 05$''$.732 (3)   & {\em d} \\
T Tau S (VLBI)        & 4$^h$  21$^m$  59$^s$.4263 (14)&  19$^{\circ}$  32$'$ 05$''$.730 (20)  &         \\   \hline 
\end{tabular}
\end{center}
\label{allpostab}
\end{table*}

\begin{figure}
\psfig{figure=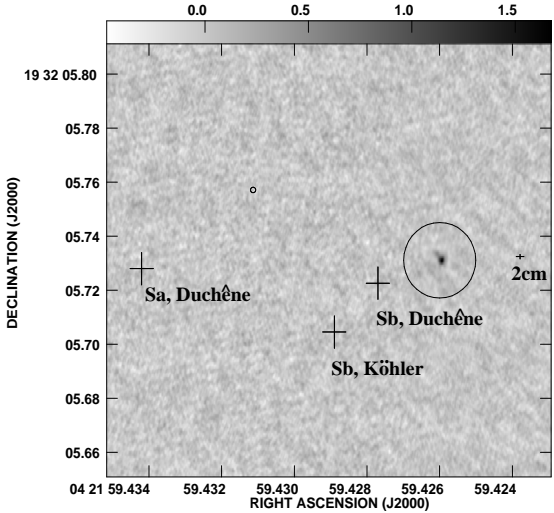,width=3.truein,height=3.truein}
\caption{The VLBI T~Tau~S field. All positions have been proper motion
corrected to the epoch of the VLBI observation but no attempt has been
made to correct for possible orbital motion of components.  A compact
radio source is clearly seen.  The circle plotted around this shows
the estimated position uncertainty of the VLBI observation at
3.6cm. The positions of T~Tau~Sa and two positions for T~Tau~Sb, one
from K\"ohler et al. and one from Duch\^ene et al., are marked with
crosses. The size of the crosses show the position uncertainties. The
2cm radio position measured by Johnston et al at epoch 2001.0531 is
shown as a small cross. Also marked (small circle) is the position
expected for Sa if the radio source is taken to be Sb.  The size
of this small circle corresponds to the relative position uncertainty
reported by K\"ohler (2000).} 
\label{posnfig}
\end{figure}

The VLBI map of the T~Tau~S field is plotted in
Figure~\ref{posnfig}. A single pointlike source can be seen at
position $\alpha=4^{h}~21^{m}~59^{s}.4263$,
$\delta=19^{\circ}~32'~5''.730$ (equinox 2000). This position is epoch
1999.958, and has been corrected for parallax using the Hipparcos
value of 5.66~mas.  The astrometric accuracy was estimated from the
position of the secondary calibrator J0357+2319 in the hybrid map
compared to its expected position from the VLBA survey. The detected
source was found to be offset 23~mas to the east and 17~mas to the
south of the catalogue position.  Assuming that this offset arises
from a combination of position errors for the primary and secondary
calibration sources, we adopt an uncertainty at the T~Tau~S position
will be $\sqrt{2}\times$~20~mas, or 14~mas. This uncertainty is
indicated with a circle around the source in Figure~\ref{posnfig}.

We can compare the position of the VLBI source to the expected
positions of T~Tau~Sa and Sb in the infrared from the recent
literature. These positions are available as offsets from the IR
position of T~Tau~N. We take as the fundamental reference position the
Hipparcos position of T~Tau~N, which is determined in the V-band.
The epoch of the Hipparcos observation was 1991.25. The Hipparcos
proper motion of $\alpha=$15.45$\pm$1.88,
$\delta=-12.48\pm1.62$~mas~yr$^{-1}$ was applied to correct the
T~Tau~N position to the epoch of our observation. The standard error
in the T~Tau~N position at the epoch of the observations is then
$\Delta \alpha=$16.4~mas, $\Delta\delta=$14.1~mas, and is dominated by
the error arising from the proper motion correction.

The IR observations of Duch\^ene et al. (2002) give positions for Sa
and Sb as offsets from the IR position of T~Tau~N at epoch 2000.2.  We
have used these offsets to calculate the positions for Sa and Sb which
are given in Table~\ref{positions}.  The orbital motion of T~Tau~S
about T~Tau~N was estimated by Roddier et al. (2000) who found that
the position angle of T~Tau~S relative to T~Tau~N changed by
$0^{\circ}.66$~yr$^{-1}$. This translates into 8~mas per year in a
westward direction. Since Sa is the brightest component in the IR (by
about 2.5 magnitudes in K), we assume that this represents motion of
T~Tau~Sa around T~Tau~N, although the motion of the photocentre could
be influenced by the motion of T~Tau~Sb around T~Tau~Sa. Since any
motion between the epoch of our observation and that of Duch\^ene et
al.  would be small, we have not attempted to account for it in the
position of Sa. This assumption is also made by Johnston et al, and
their observations are consistent with orbital motion of T~Tau~Sb
around a stationary (with respect to T ~Tau~N) T~Tau~Sa.  The
observation of the T~Tau~S system by K\"ohler et al. (2000) is closest
in time to our observation (epoch 2000.17 as opposed to 1999.96), and
so this is the position for T~Tau~Sb with which the VLBI source
position should be principally compared.

Finally, in Table~\ref{positions} we list the position of the southern
2cm source observed by Johnston et al. at epoch 2001.0531, which has
been proper motion corrected to epoch 1999.958 using the proper motion
reported for T~Tau~S by these authors.

All the positions are listed in Table~\ref{positions} and are shown in
Figure~\ref{posnfig}, with the exception of T~Tau~N which lies outside
the field.  The T~Tau~Sa and Sb positions are marked with crosses,
the sizes of which indicate the uncertainty in the positions which are
dominated by the proper motion correction to the T~Tau~N position as
previously noted. This component of the uncertainty is the
same for Sa and Sb, and so the uncertainties shown for these sources
are correlated.

The VLBI source in our map is closest to the position of T~Tau~Sb, and
we make this identification, although there is a significant offset
from the K\"ohler position. One way this might arise is because the
T~Tau~N position is from the {\em V}-band Hipparcos observations,
whereas the position of T~Tau~Sa relative to T~Tau~N by Duch\^ene et
al. was determined in the near infrared. The presence of scattered
light in the NIR may cause a discrepancy between the optical and NIR
positions which then leads to a discrepancy between the T~Tau~S
positions and the radio position.  Possible orbital motion of the
T~Tau~Sa+Sb system in the interval between our observations at epoch
1999.958 and the determination of the T~Tau~Sa position relative to
T~Tau~N at epoch 2000.9 can be ruled out as an explanation, since the
motion is both too small and in the wrong sense to account for the
discrepancy.

We also show in Figure~\ref{posnfig} the position expected for a radio
source which could be identified as T~Tau~Sa, based on the hypothesis that the
detected radio source is in fact T~Tau~Sb. This position is again
determined from the K\"ohler et al. separation and position angle relative to
the Sb position, and is marked with a small circle, the radius of
which corresponds to the uncertainty in K\"ohler et al.'s position
(there is no contribution here from the uncertainty in the T~Tau~N
proper motion correction).  It is clear that no radio source appears
in the map at or near this position. We have also searched in the
opposite position for a source corresponding to T~Tau~Sb based on the
hypothesis that our detected source is in fact Sa, and this search was
also unsuccessful.

A search was made in the VLBI map for other components of the system,
namely T~Tau~N and the hypothetical source labeled T~Tau~R by Ray et
al. (1997). As Johnston et al. (2003) we did not detect the supposed
T~Tau~R component.  The expected region of T~Tau~N lies outside the
primary beam of the phased VLA, and so the search for this object
lacks the most sensitive antenna. Phase problems may also degrade the
image at such a large distance from the correlation centre. Despite
the favorable RMS in the map (0.1\,mJy), no sources were found
anywhere in the region of T~Tau~N. No extended background flux was
seen either. We conclude that the majority of the T~Tau~N flux is
distributed over an area larger than about 20~mas, the resolution
obtained at the shortest baseline available (Los Alamos-Pie Town,
about 4M$\lambda$, see section~\ref{sizesection}).

\begin{figure*}
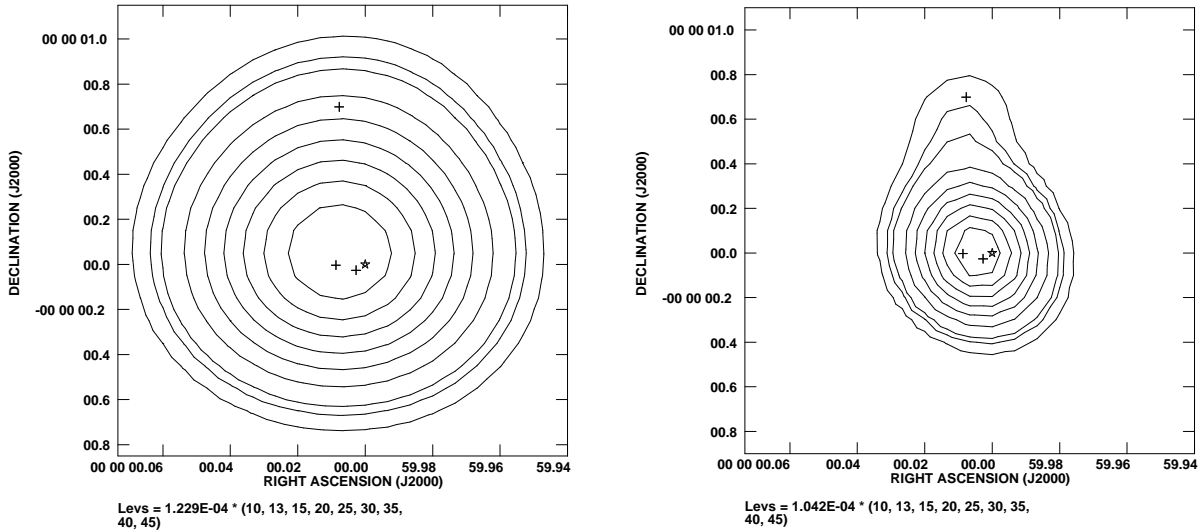

\hbox{
\psfig{figure=Fig2a.ps,width=3.truein,height=3.truein}
\hspace{0.5cm}
\psfig{figure=Fig2b.ps,width=3.truein,height=3.truein} 
}
\caption{Contour maps of the T~Tau field from the VLA only data. The
map on the left has been cleaned using the fitted beam of
1.15$\times$0.94$''$, whilst the map on the right has been cleaned
using a restoring beam set to $0.5\times0.5''$ to produce an
artificial overresolution of the data. This allows T~Tau~N to be
distinguished from T~Tau~S for the purposes of flux
measurement. Positions are given as offsets from the position of the
VLBI source, which is marked with a star.  The position of T~Tau~N,
T~Tau~Sa and T~Tau~Sb (from Duch\^ene et al., epoch 2000.2) are marked
with crosses.}  
\label{vlamap} 
\end{figure*}

\subsection{Fluxes and time variability}

\begin{figure*}
\vbox{
\hbox{
\psfig{figure=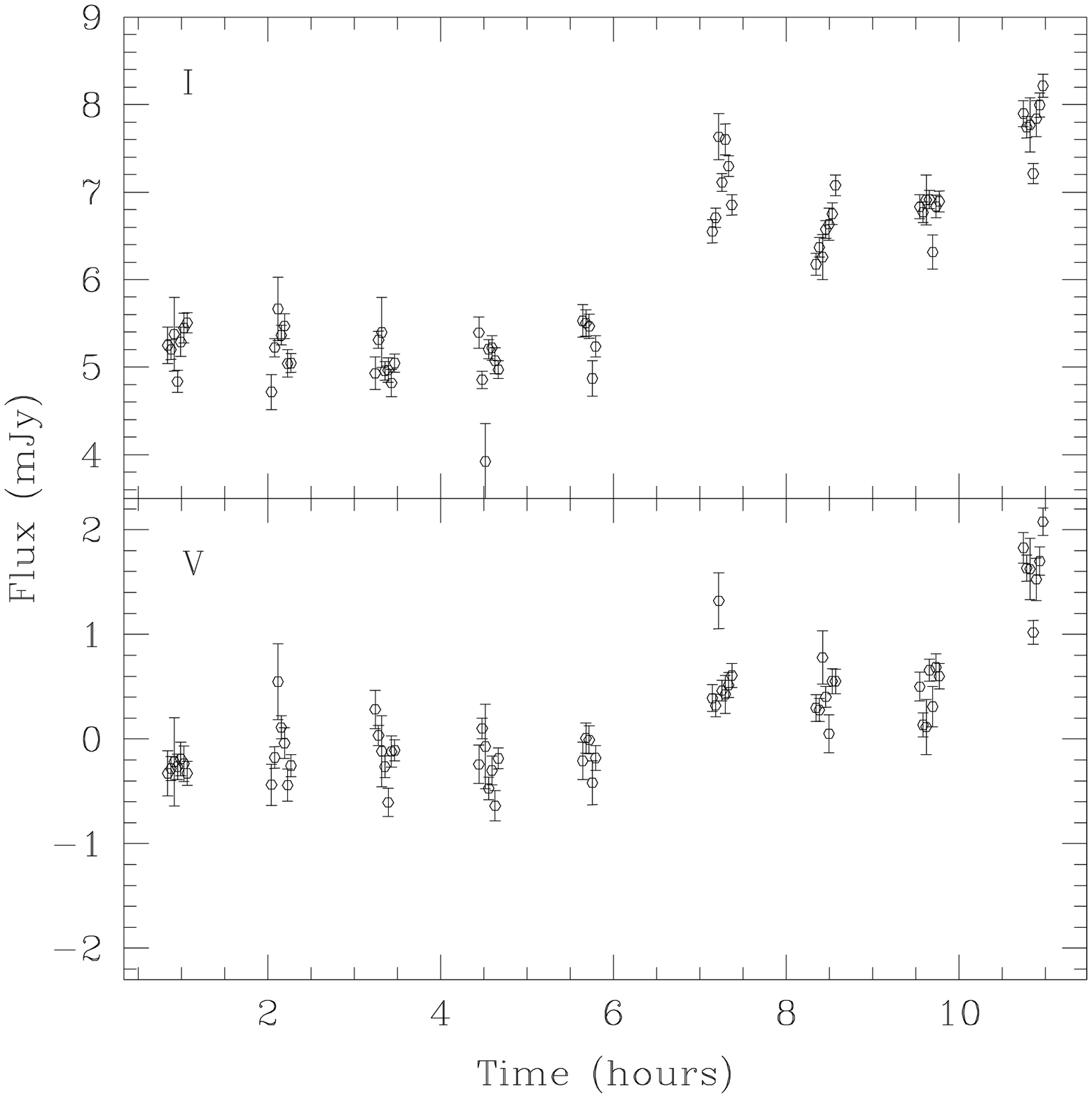,width=3.truein,height=3.truein}
\psfig{figure=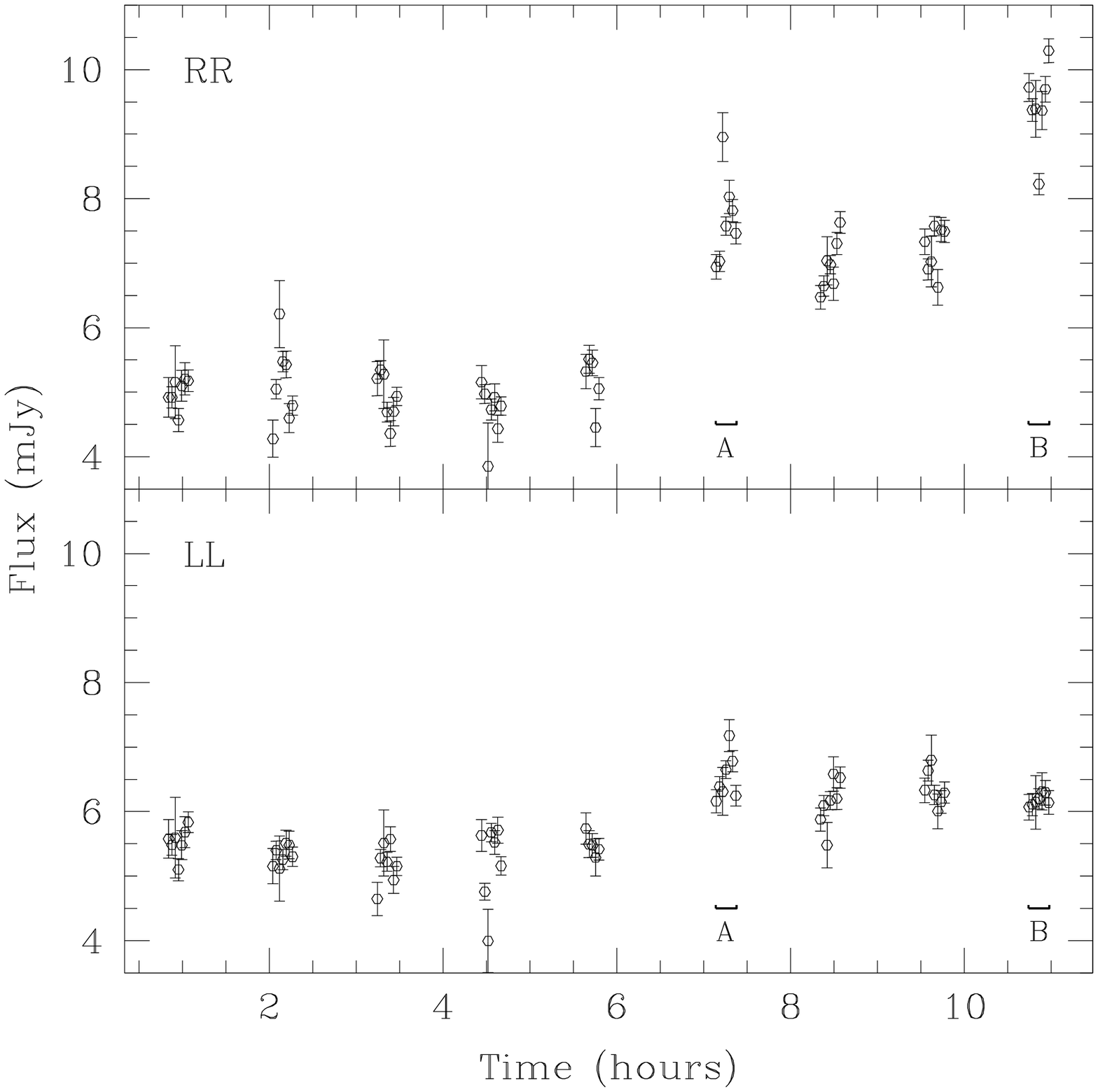,width=3.truein,height=3.truein}
}
\hbox{
\psfig{figure=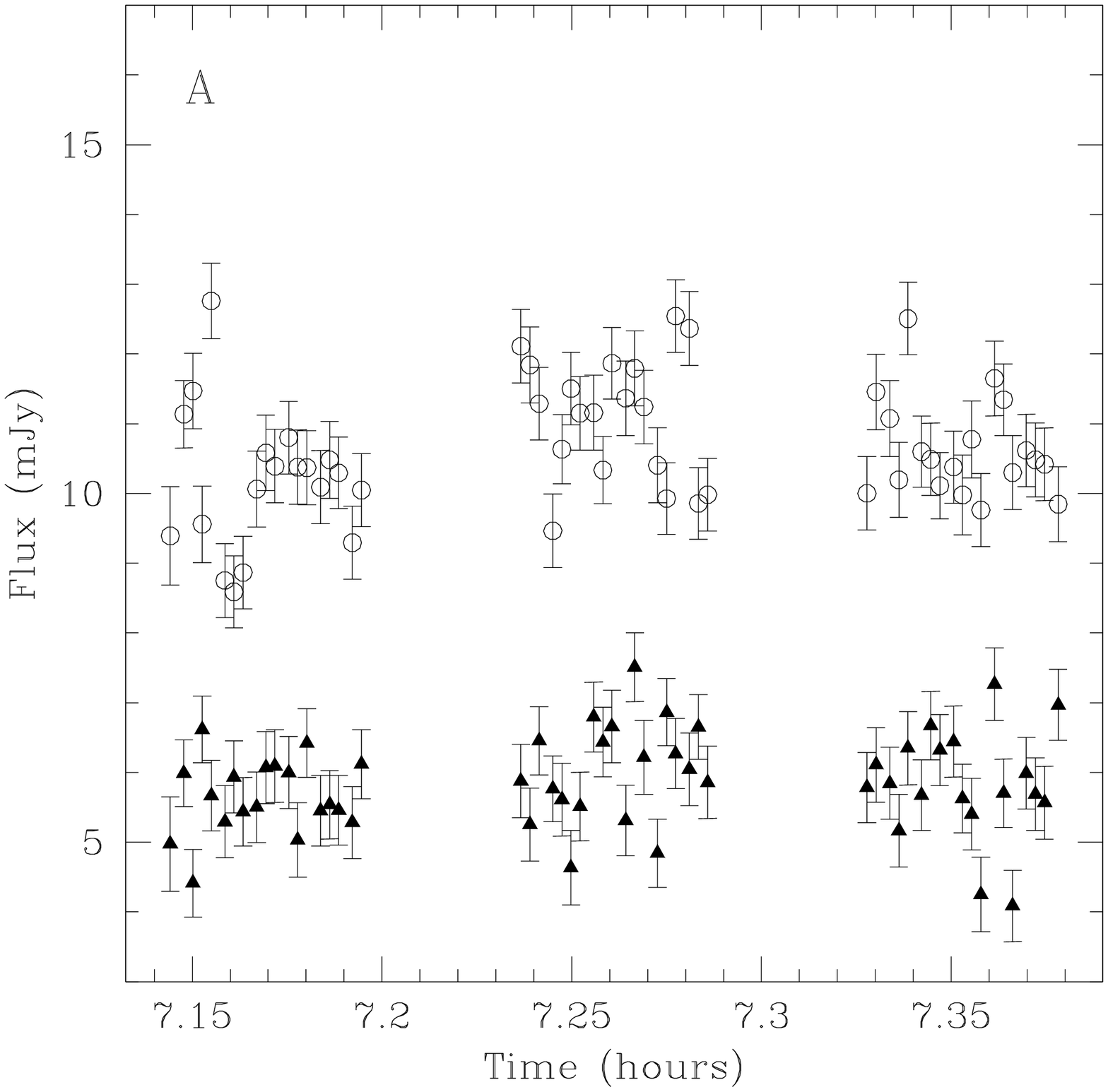, width=3.truein,height=3.truein}
\psfig{figure=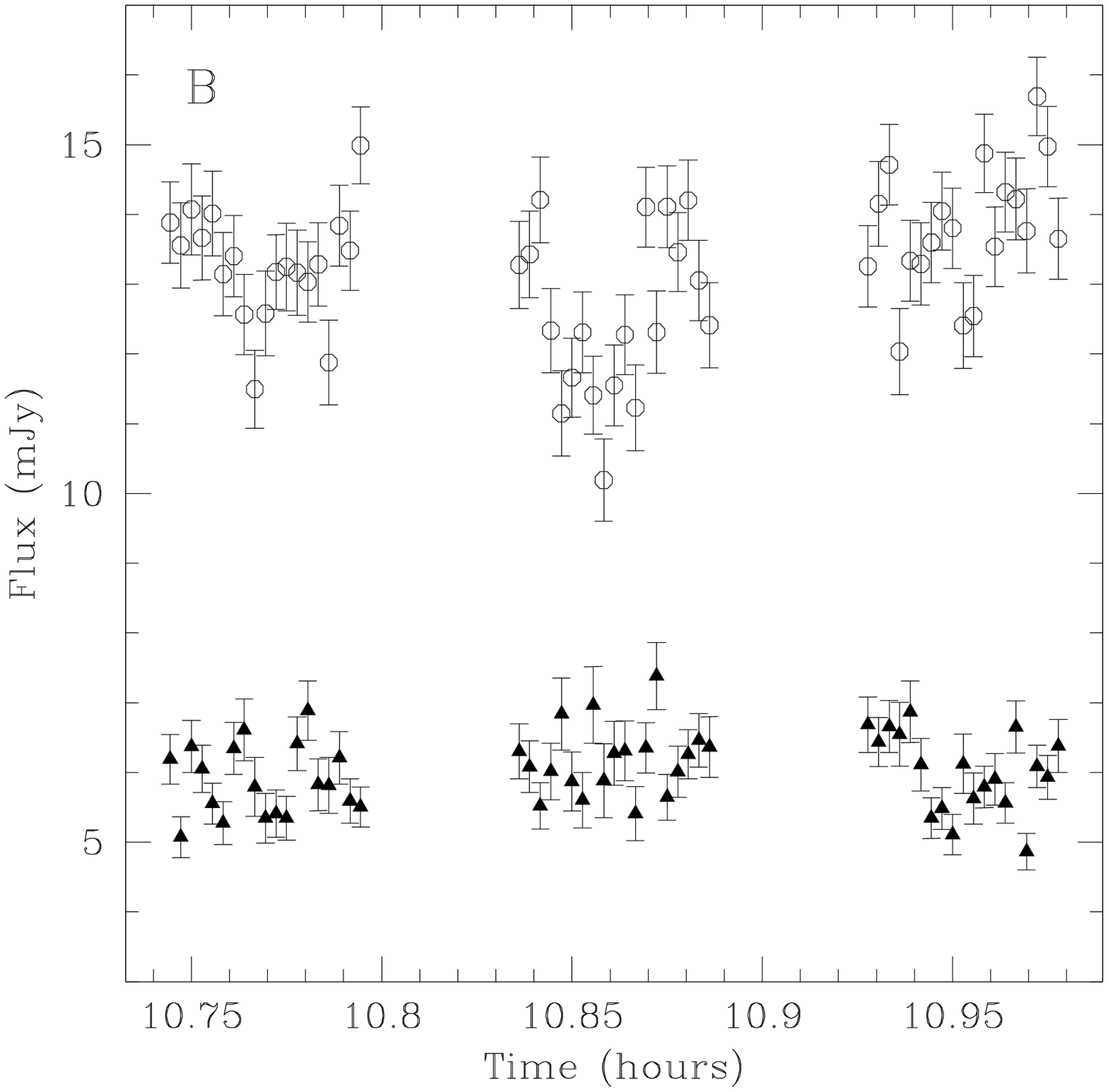, width=3.truein,height=3.truein}
}
}
\caption{Upper panels: The VLA lightcurves. In the left panel are
shown Stokes I and V, and in the right Stokes RR and LL.  Every group
of points includes three on-source scans.  Two sharp increases in flux
are seen.  Both show significant right-hand circular polarization, and
the second appears to be almost 100\% polarized.  Lower panels: Blow-ups 
of two sections of special interest, showing rapidly varying
behaviour.  These sections are indicated in the upper right-hand panel.  
Stokes RR and LL are plotted separately.
The RR points are plotted as open circles, the LL points as filled
triangles. The RR points in both figures have been shifted upwards by
4~mJy to prevent overlap.  Successive points are approximately 10~s
apart, which corresponds to the basic integration time and is
therefore the highest time resolution available in our data.}
\label{ttsvlalc}
\end{figure*}

The VLA lightcurve is shown in Figure~\ref{ttsvlalc}. The source
brightened considerably during the observing run. Two large increases
in flux occurred, one at around UT=7.15 hours and one at around 
UT=10.75 hours.  Flaring behaviour can be seen at the beginning of the
scan commencing at UT=7.14 hours. The
upper right-hand panel of Figure~\ref{ttsvlalc} shows RCP and LCP behaviour
separately. The polarization of the overall flux changes from about
5\% LCP to about 6\% RCP. The flare from UT=7.15 to 7.16 hours is
almost totally right-hand polarized.
This is shown at higher time
resolution in the lower left-hand panel of Figure~\ref{ttsvlalc}.

After UT=6~hours there is a long period of increased emission, with a
slow decay, lasting until the second flux increase which is seen in
the lightcurve beginning at UT=10.74 hours. Here, we did not see any
sign of an initial flare within a scan (Figure~\ref{ttsvlalc},
lower right-hand panel), although ongoing short-timescale variability can be
seen in the RR flux, which is not observed at LL and which is greater
than any increased scatter expected from Poisson statistics due to the
higher flux level. The flux level at UT=10.74 hours persisted until
the end of the observation at UT=11~hours, with no sign of decay.  The
extra flux appearing after UT=10.74~hours is 100\% right hand
polarized. In fact, the lower panel of the right hand side of
Figure~\ref{ttsvlalc} shows that the LL flux continues its slow
decline from the previous increase at UT=7.15~hours. The overall
polarization during the final three scans was $\sim 20\%$ RCP.

The VLA in `B' configuration results in a beam of approximately $1''$
FWHM. This is somewhat larger than the separation of T~Tau~N and
S. Nevertheless elongation could be seen in the maps.  The radio
emission is dominated by the southern component. The peak flux in the
VLA map is 5.2$\pm$0.02 mJy. By setting the restoring beam to be
smaller than the measured beamsize, the north and south components can
be more clearly distinguished. We chose a restoring beam of 0.5$''
\times 0.5''$.  This allows the measurement of fluxes for T~Tau~N and
T~Tau~S separately (see Table~\ref{tabfluxes}).  The VLA map with
fitted restoring beam and artificially small restoring beam are shown
in Figure~\ref{vlamap}. Both peak and integrated flux densities were
determined from these maps within a box containing each object. The
peak flux of T~Tau~S increased from 4.5\,mJy before to 6.1\,mJy after
the first outburst and then to 7.1\,mJy after the second, whilst
T~Tau~N remained steady at approximately 1.8\,mJy.  Since the
instrument beam is of comparable size to the source separation, there
will inevitably be some cross-contamination in the flux density
measurement, despite the small restoring beam size.  The effect of
this can be seen in the increasing flux density measured for T~Tau~N
as the southern source brightens. The steady peak flux for T~Tau~N
suggests that this rise in the flux density is driven by contamination
from T~Tau~S, and that the variability is dominated by the southern
source.

The difference between the peak and integrated flux is a crude
estimate of the resolved flux at the VLA. It appears that T~Tau~N is
pointlike at this resolution, but T~Tau~S may have an extended
component in the VLA maps (column 5 of Table~\ref{fluxes}). 
Likewise, the difference between the integrated
VLA flux and the VLBI flux can be used to estimate the flux which is seen 
by the VLA but resolved out at the shortest VLBI baselines.
Here, an extended component associated with T~Tau~S has a flux of
around 4.5\,mJy and its size can be estimated to be 100-300~mas.  The
southern source therefore comprises a compact source, unresolved by
the VLBI, plus extended emission which is resolved at VLA baselines.

\begin{table*}
\caption{\label{fluxes} Comparison of fluxes (in mJy) between VLA and VLBI
maps. The peak and integrated flux values for each object in each time
segment (0-6 hours, 6-10 hours, 10-11 hours) are listed for the VLA
maps. The VLBI fluxes are measured from the visibility 
plots shown in Figure~\ref{uvplt1}. All values are in mJy, with uncertainties in
parantheses. The difference between the VLA integrated flux density and the
VLA peak flux indicates the resolved flux in the VLA map (5th
column). The difference between VLA peak and VLBI fluxes indicates the
amount of flux which is compact at the VLA but resolved out by the
VLBI (6th column). }
\begin{center}
\begin{tabular}{|c|c|c|c|c|c|} \hline
\multicolumn{6}{|c|}{T Tau S} \\ \hline
Time          & VLA peak flux & VLA Integrated flux    & VLBI flux       & VLA(integrated)-VLA(peak) & VLA(integrated)-VLBI  \\ \hline
0 to 6 hr     &  4.46 (0.04)        & 5.97 (0.44)      & 1.4 (0.1)       & 1.51 (0.44)                 &   4.57 (0.5)          \\ 
6 to 10 hr    &  6.13 (0.04)        & 7.46 (0.44)      & 3.3 (0.2)       & 1.33 (0.44)                 &   4.16 (0.5)          \\
10 to 12 hr   &  7.06 (0.1)         & 8.81 (1.1)       & 3.8 (0.4)       & 1.75 (1.1)                  &   5.01 (1.2)       \\  \hline     
\multicolumn{6}{|c|}{T Tau N} \\ \hline
              &               &                        &                 &                       &                       \\ \hline
0 to 6 hr     &  1.75 (0.04)        & 1.67 (0.3)       & --              & -0.08 (0.3)           &   --           \\ 
6 to 10 hr    &  1.77 (0.04)        & 1.80 (0.3)       & --              &  0.03 (0.3)           &   --            \\
10 to 12 hr   &  1.79 (0.1)         & 2.11 (0.7)       & --              &  0.32 (0.7)           &   --             \\  \hline 
\end{tabular}
\end{center}
\label{tabfluxes}
\end{table*}

\subsection{The size of the VLBI source}
\label{sizesection}

Due to the large variations in flux, the data were
analysed separately in three time segments, 0 to
6 UT, 6 to 10 UT and after 10 UT.

\begin{figure*}
\psfig{figure=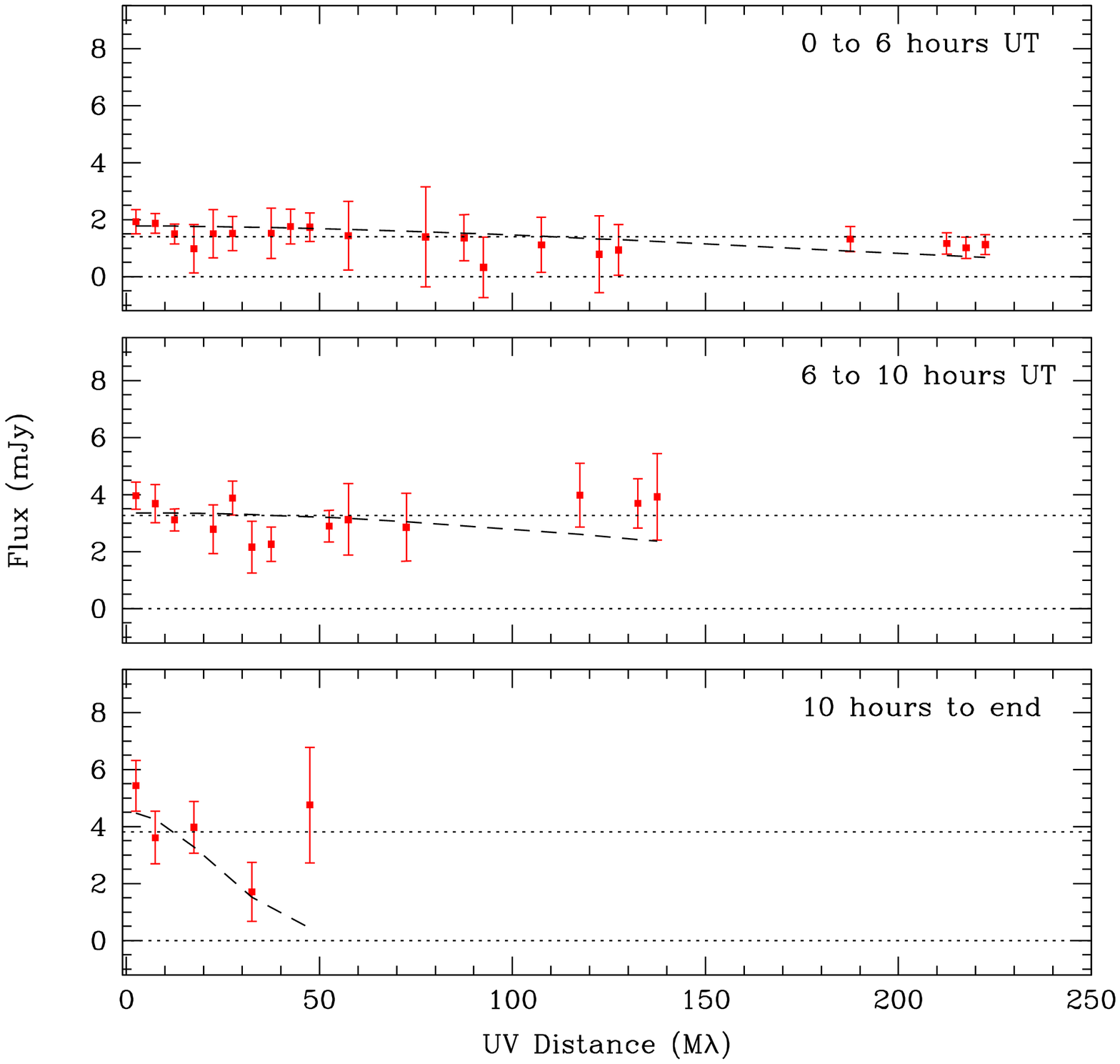,width=6.truein,height=6.truein}
\caption{Correlated flux vs {\em uv} distance. The panels show (top to
bottom) the visibility amplitude for the time periods 0 to 6 hours, 6
hours to 10 hours and 10 hours to the end of the observation. These
time segments correspond to the three stages of flaring behaviour seen
in Figure~\ref{ttsvlalc}. The dotted straight lines mark the weighted
data average. Dashed lines show the narrowest possible gaussian
fitting the data. See the discussion in the text for details of the
fits.}
\label{uvplt1}
\end{figure*}

\begin{figure*}
\hbox{
\psfig{figure=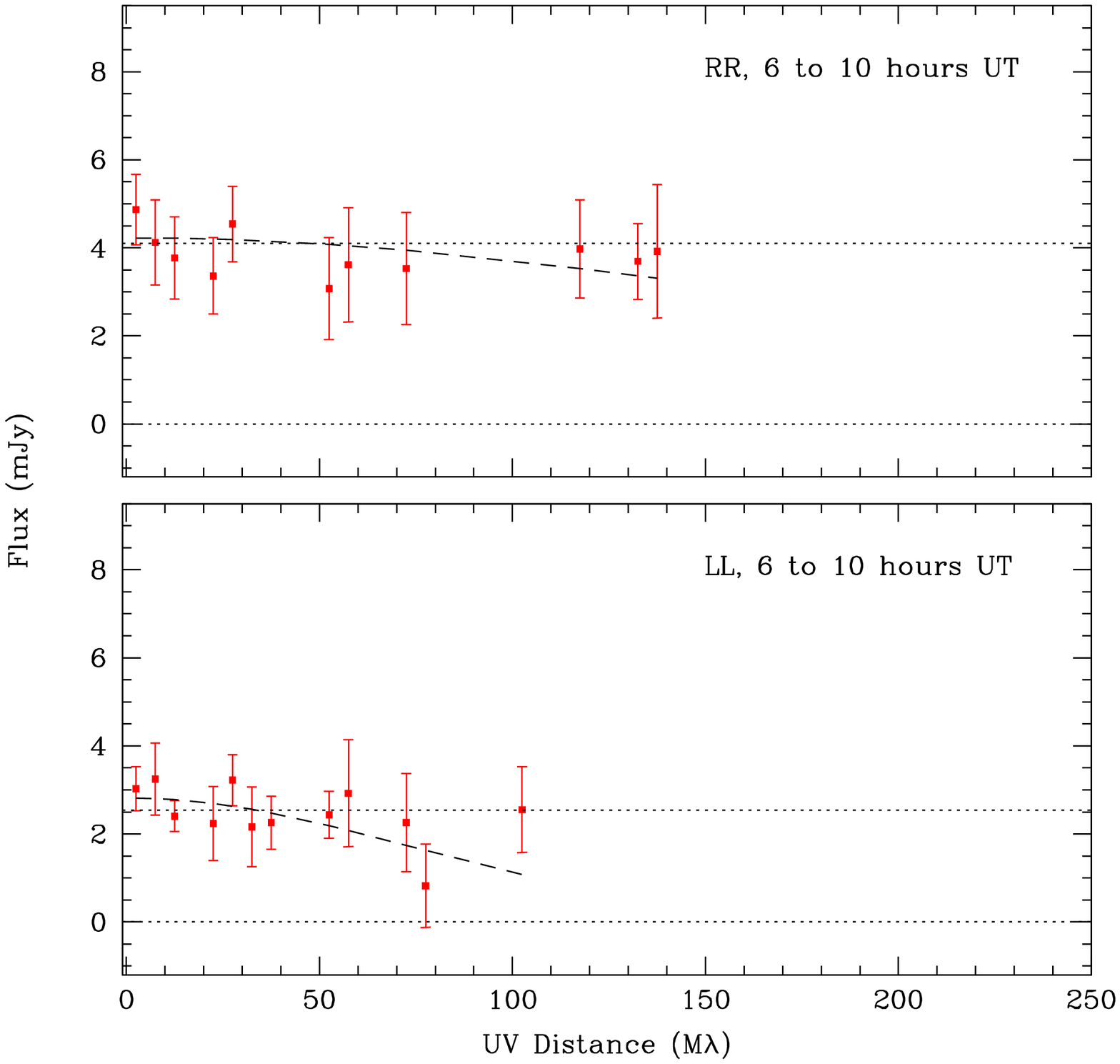,width=3.truein,height=3.truein}
\psfig{figure=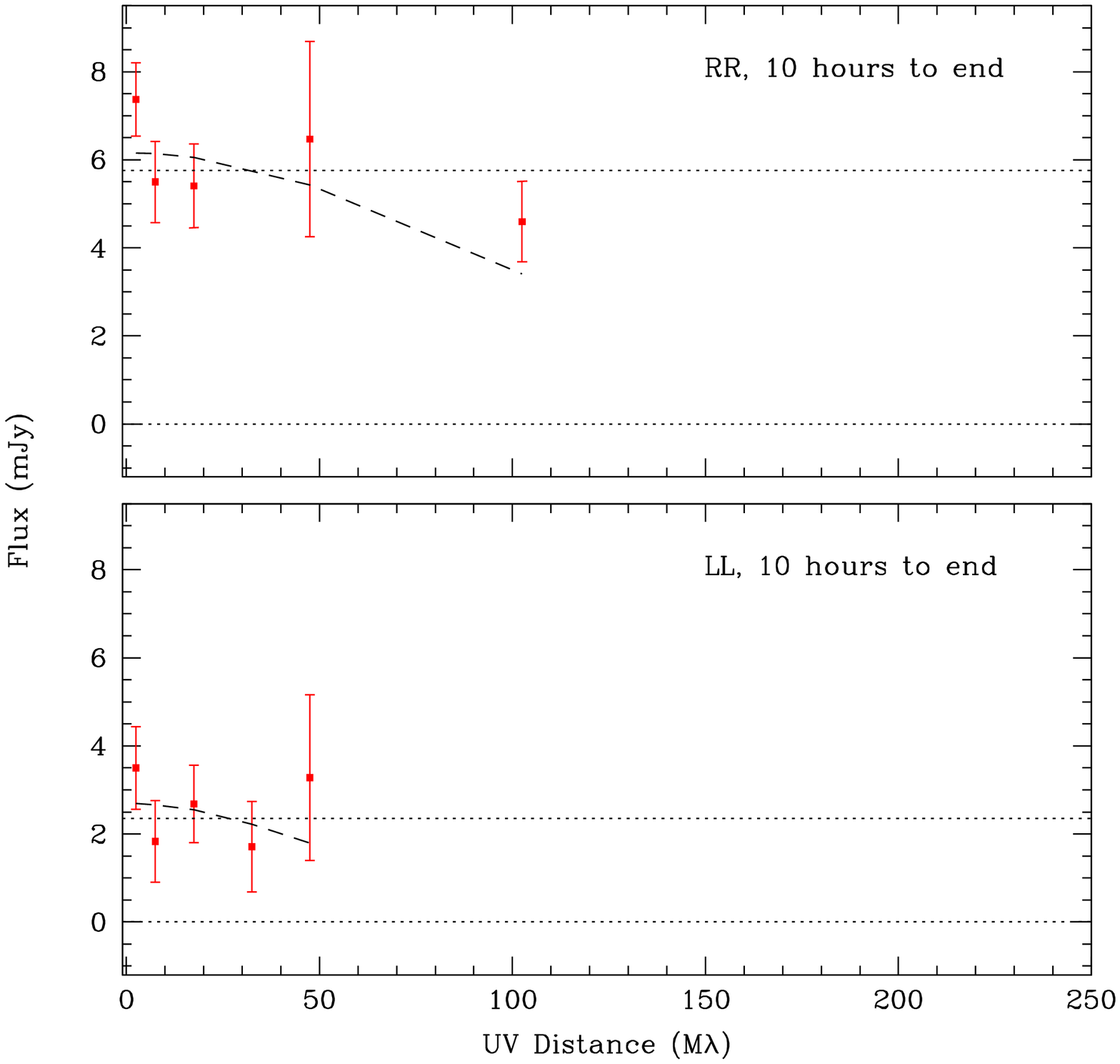,width=3.truein,height=3.truein}
}
\caption{Correlated polarized flux vs {\em uv} distance for the periods of
the two flares, separated in RR and LL. For the description of the
plots see Figure \ref{uvplt1}. The LL data for Mauna Kea after UT=10
hours, which forms the longest baselines at this time, was flagged as
unreliable and so the long baseline point is missing in the lower
right-hand panel.}
\label{uvplt2}
\end{figure*}

Plots of correlated flux against baseline distance for baselines
including the VLA are shown in Figures~\ref{uvplt1} and~\ref{uvplt2}.
These plots were produced in the following way.  For each baseline the
complex phase-referenced visibility data were first averaged
coherently in time over 25 - 30 minutes (the length of the scan groups
shown in Figure~\ref{ttsvlalc}), and then coherently averaged over the
two IF channels, eight frequency channels and the RR and LL
polarizations.  From the scatter of the data over time error bars were
assigned. These were propagated through subsequent averaging steps to
determine the final error bars. On many baselines the phase versus
time variations could be detected within the scans and be seen to be
much less than a radian. Based on this and on the fact that the total
flux in our VLBI map is comparable to the peak flux, we do not believe
this averaging produced any significant loss of amplitude due to
coherence losses over the averaging time chosen. At this point all
baseline visibilities had signal to noise ratio significantly greater
than unity.  Finally the baseline amplitudes within annular shells in
the {\em uv} plane were incoherently averaged and the results plotted as a
function of baseline length. Because the signal to noise ratio of the
original baseline averages were all greater than one the effect of
noise-bias on the incoherently averaged amplitudes shown in
Figure~\ref{uvplt1} is small. Given this, symmetric error bars are
plotted.

The source is detected during the first interval from 0 to
6 hours with a flux around 1.4~mJy over the short baselines and is
more clearly detected at the longest, but much more sensitive,
Effelsberg baselines. The steady brightening in the following two
periods can be clearly seen. It is unfortunate that Effelsberg
baselines were only available for the first four hours of the
experiment, so any information associated with the structure of the
flaring source on such small scales is unavailable
($\sim10$R$_{\odot}$).

We show in each plot a line marking the weighted average of the data
points, which corresponds to a point-source model. In all cases the
straight line is an acceptable fit to the data ($\chi^2=0.42$ for
0-6~hours, $\chi^2=0.75$ for 6-10 hours and $\chi^2=1.6$ for
10-11~hours).  However, we wish to determine a meaningful upper limit
for the source size.  We have done this by fitting the narrowest
possible Gaussian (representing the largest possible source) with the
same $\chi^2$ as the straight line.  The Gaussian amplitude is left as
a free parameter. In the second data segment (6 to 10 hours UT), the
$\chi^2$ for the fit never converged to the straight line. We
therefore show the narrowest fit which attained $\chi^2=1$. The values
of HWHM for the fitted Gaussians and HWHM of the corresponding
Gaussian source on the sky are given in Table~\ref{srcsizes}.

In Figure~\ref{uvplt2} the right- and left-handed polarizations are
shown separately for the time segments 
6-10 hours and 10 hours to the end of the experiment.
The Stokes LL data 
from Mauna Kea were flagged during the last part of the 
experiment, and for this reason the longest baseline is missing from the 
final LL plot (lower right-hand panel). The enhanced brightening at
RR can be clearly seen. Also clearly seen is the fact that LL does not
brighten at all during the second flare, whereas RR does.  This is
consistent with the lightcurve shown in Figure~\ref{ttsvlalc}. We have
determined upper limits to the source sizes for the separate RR and LL
data in the same way as described above. These are also listed in
Table~\ref{srcsizes}.

\section{Discussion}

\subsection{T Tau N}

T~Tau~N is not resolved in the VLA map, which has a maximum baseline
separation of $\sim$300~k$\lambda$, but is resolved-out by the
VLBI. This constrains its size to be between roughly 20~mas and
300~mas. A 20~mas source corresponds to 2.8 AU at 140~pc, which is
large enough to rule out a coronal or magnetospheric origin for the
flux and instead point to a wind source. The lack of variability or
polarization tends to reinforce this conclusion.

Using the equations given in Panagia \& Felli (1975) for a fully
ionized, isotropic, optically thick envelope with $n_e \propto
r^{-2}$, and assuming the outflow velocity to be 240kms$^{-1}$ (after
Skinner \& Brown) we can estimate the mass loss rate to be $5\times
10^{-8}$~M$_{\odot}$yr$^{-1}$ for a source of size 20~mas or
$8\times10^{-8}$~M$_{\odot}$yr$^{-1}$ for a source of size
500~mas. These values are up to a factor of 2 higher than those
estimated by Skinner \& Brown (1994). Because we have constraints both
on the source size and the flux, we can also estimate the electron
temperature. For a source of size 20~mas, the electron temperature
would be around $2\times10^{4}$~K. For a source of around 80~mas, the
implied electron temperature would fall to around 5000K. Sizes larger
than this are therefore unlikely.  The corresponding brightness
temperatures can be calculated from
\begin{equation}
T_b = \frac{c^2 S_{L,R}}{\nu^2 k \Delta \Omega} = 4.43\times10^8 \nu_9^{-2} F_{mJy} r_{mas}^{-2},
\end{equation} 
where $S_{L,R}$ is the Planck function, $\nu_9$ is the frequency in
GHz, $F$ is the total source flux (including both polarizations) 
and $r$ is the source size (HWHM).  For an 80~mas
source, $T_b=7055~K$. For 20~mas, it is $1.1\times10^{5}$~K.  These
values are consistent with the interpretation of thermal wind
emission.

\subsection{T Tau Sa}

No source could be seen corresponding to T~Tau~Sa. The RMS noise in
the region of the map where Sa was expected to lie was approximately
0.1~mJy. A comparison of the flux at the peak of the VLA map with the
flux of the compact source identified as Sb reveals that around 4.5~mJy
can be attributed to a source which appears compact to the VLA
($\theta < 500$~mas), but is apparently resolved-out by the VLBI array
($\theta > 50$~mas). Only a small fraction of this flux will be
contamination from T~Tau~N, some may correspond to the extended
structure seen by Ray et al., and some may be due to an extended source
associated with T~Tau~Sa.

\begin{table*}
\caption{\label{srcsizes} Upper limits on source sizes at various
times for different polarizations. The size is given as the half width
half maximum of the fitted Gaussian, $H$, the corresponding HWHM on the sky in
milliarcseconds and the physical source size for a distance of 140pc (Elias,
1978) in solar radii. }
\begin{center}
\begin{tabular}{|l|ccc|ccc|ccc|} \hline
        &                  \multicolumn{9}{c|}{Polarization}                                          \\ \hline
        & \multicolumn{3}{c}{I} & \multicolumn{3}{c}{RR}   &  \multicolumn{3}{c|}{LL}                \\ \hline
Time    & H (M$\lambda$)  &  r (mas)& (R$_{\odot}$) & H (M$\lambda$)  &  r (mas) & (R$_{\odot}$) & H (M$\lambda$)  &  r (mas) & (R$_{\odot}$) \\ \hline 
0 - 6   &  189               &  0.48   & 14.4       & 172             &  0.53    &  15.9         & 192             &  0.47     & 14.3 \\
6 - 10  &  193$^*$           &  0.47   & 14.2       & 231             &  0.39    &  11.9         & 87              &  1.05     & 31.5 \\  
10 - 12 &  26                &  3.5    & 105.4      & 111             &  0.81    &  24.7         & 62              &  1.47     & 44.2 \\ \hline
\end{tabular}
\end{center}
\end{table*}

\subsection{T Tau Sb}

The flux of T~Tau~S comprises a steady, extended component of about
4.5~mJy, which seems to be connected either to the circumbinary
environment or to T~Tau~Sa, together with a compact, varying component
which can be identified as T~Tau~Sb.  This has a quiescent level of
1.4~mJy rising up to steady levels of 3.8mJy, with some spikes up to
around 7~mJy.  The {\em extra} flux in the two brightenings in the VLA
lightcurve at UT=7 hours and UT=10hours is consistent with the extra
flux seen in the VLBI data segments.  This shows that the compact VLBI
object can be confidently identified as the source of the rapidly
varying flux. We found in the VLBI map no sign of the extended
polarized structures reported by Ray et al. (1997) tracing the
outflow, or of any structure to the northeast which could correspond
to that seen by White and coworkers (White, 2000). The beam size in
the Ray et al. maps was of the order of 150~mas, the individual left-
and right-hand polarized blobs appear pointlike at this resolution in
their maps, and the offset between them was around 200~mas.  The
structure seen by Ray et al. would not be expected to be resolved in
low resolution VLA map, but could nevertheless be resolved out in the
VLBI data. The flux of 2.1~mJy reported by Ray et al. is below the
flux level of around 3~mJy which we believe is resolved out by the
VLBI (see Table~\ref{fluxes}).  A source size of around 100~mas
corresponds to a scale around 1M$\lambda$ in the visibility
functions. The shortest VLBI baseline is VLA-Pie Town (PT), with a
{\em uv} distance of around 1M$\lambda$. We see no evidence of
resolution within the set of VLA-PT scans, or between these and the
next shortest baseline. Nevertheless, our visibility amplitudes are
poorly sampled on these scales, and we cannot rule out that there is
structure there which is missed. It should of course be borne in mind
that these sources are intrinsically variable. Furthermore, in the
case of blobs of outflowing material, significant motion may well have
occurred on time scales of years between the observations.

The circular polarization is an indicator of the presence of magnetic
fields. The short timescale variability, from hours down to seconds,
together with MK brightness temperatures, points to a gyrosynchrotron
emission mechanism.  The 100\% right-hand polarized emission which is
seen in the spike at 7 hours UT and again more prominently at 11 hours
UT probably arises from a coherent emission mechanism. This is
discussed further below.

The initial luminosity was around
$1.3\times10^{17}$~erg~s$^{-1}$~Hz$^{-1}$. The extra flux appearing
after UT=7 hours has a luminosity of
$4.4\times10^{16}$~erg~s$^{-1}$~Hz$^{-1}$. The luminosity of the extra
flux appearing after UT=11 hours was
$1.8\times10^{16}$~erg~s$^{-1}$~Hz$^{-1}$. 
Both the steady luminosity and the extra flux appearing during the 
variations have similar values to those observed in typical RS CVn 
systems, whose flux is also believed to arise in large-scale 
powerful magnetic fields (e.g. Benz \& G\"udel, 1994). 
 
The polarization flip which was observed with the first flux increase
at UT=7 hours is similar to that seen by Skinner \& Brown (1994).
They attributed this to a change in the relative optical depth of the
$o$- and $x$-modes during the brighter phase, after a model for RS CVn
systems developed by Morris, Mutel \& Su (1990). Since these modes
correspond (under normal conditions) to different observed
polarizations, such changes can lead to the observed polarization
flip. Another possible explanation would be a change in the
sense of the dominant magnetic field along the line of sight, perhaps caused
by stellar rotation, or the addition of new highly-polarized emission.

The size of the compact source is constrained from the VLBI data to be
less than 0.48~mas. At a distance of 140~pc this corresponds to a
scale of 14.5~R$_{\odot}$. This length scale is not much larger than
the expected size of a T~Tauri magnetosphere. We can also estimate the
size of the varying region from the timescale of the flux variations.
The most rapid variations take place on timescales down to our
fundamental integration time of 10~s. 
This indicates a source of length scale at most $c\times dt=
3\times10^{9}$~m, or 4.3~R$_{\odot}$. This yields a lower limit to the
brightness temperature of around $2\times10^9$~K for the
region involved in the rapidly varying emission in each case.  Whilst
this is still consistent with gyrosynchrotron emission from
non-thermal electrons, the 100\% right-hand polarization of the extra
flux seen after the second flux increase is not. A coherent emission
mechanism must be invoked to produce this strong polarization.

The most likely coherent emission process is an electron cyclotron maser,
arising from mildly relativistic electrons trapped in magnetic flux
tubes, where a loss-cone distribution may develop.  This emission
would occur at the fundamental or a low harmonic of the
gyrofrequency
\begin{equation}
\nu_c = \frac{eB}{2 \pi m_e c} \approx 2.8\times 10^6 B.
\end{equation}
8.4GHz emission at the fundamental or first harmonic therefore implies
a magnetic field strength $B \sim 1.5-3$kG.  Although the emission
from an individual maser pulse would typically have a frequency range
of about 1\%, the emission arises from a range of locations in the
flux tube and hence a range of magnetic field strengths, which gives
rise to a much broader spectrum.

The main alternative to this would be 
a plasma maser, with the emitting frequency being the fundamental or 
first harmonic of the plasma frequency
\begin{equation}
\nu_p = \left( \frac{n_e e^2}{\pi m_e} \right)^{\frac{1}{2}} \approx 9000 n_e^{0.5}
\end{equation}
or at the hybrid frequency $(\nu_p^2 + \nu_c^2)^{1/2}$ with $\nu_p >
\nu_c$. This would imply $n_e \sim 10^{12}$cm$^{-3}$.  

For an accreting T~Tauri
star, a possible hypothetical emitting region is a magnetically
confined accretion funnel, which would provide a region with magnetic
field, high density and high density gradients at the stream edge.
For a typical T~Tauri accretion rate of $10^{-7} - 10^{-8}$M$_{\odot}$
yr$^{-1}$ assuming the stream footpoint covers 10\% of the stellar
surface and that the inflow velocity is 150kms$^{-1}$, the density of
material in the inner stream can be expected to reach perhaps
$7\times10^{-11}$g cm$^{-3}$, or about $5 \times 10^{13}$cm$^{-3}$, 
consistent with the required density estimated above. We note that 
X-ray observations of He-like triplets in the X-ray spectrum of  
the nearby accreting T Tauri star TW Hya also indicates 
densities of the order $10^{13}$cm$^{-1}$ in the line-forming region, 
(Kastner et al., 2002).

Since for free-free absorption the optical depth for plasma emission
is $\tau \propto \nu^2$, such emission would be expected to be heavily
absorbed. A density scale length of less than 100~km would be
necessary for this process to account for the emission, assuming
second harmonic (Benz, 2002, eq. 11.2.5). The source size would have
to be the same scale or even smaller and a brightness temperature in
excess of $10^{18}$K is required. This exceeds solar radio bursts by
three orders of magnitude.

Lower limits to the brightness temperature at various times were
estimated for the upper limit source size and are shown in
Table~\ref{temps}. A second estimate was also made assuming the upper
limit source size is a constant 0.48~mas, the maximum size for the
quiescent source between the beginning of the observation and UT=6
hours.  This constraint was preferred because it is the most reliable,
including the long Effelsberg-VLA baselines, and it is considered unlikely that
the source would increase in size. The larger upper limits for the
subsequent data segments are clearly driven by the lack of the long
VLA-Effelsberg baselines. The brightness temperatures are at the high
end of the range typically seen in solar gyrosynchrotron flares
($10^7$ -- $10^8$K).

\begin{table}
\caption{\label{temps} Brightness temperature lower limits for
T~Tau~Sb. The fourth column contains brightness temperatures for the
source size constraint listed in column 2. The fifth column shows
values calculated for an assumed constant source size upper limit of
0.48~mas. }
\begin{center}
\begin{tabular}{|l|c|c|c|c|} \hline
Time    &  r (mas)           & Flux (mJy)  &  T$_b^1$ (K)     & T$_b^2$ (K) \\ \hline
0 - 6   &  0.48              & 1.4         &  3.8$\times10^7$ & 3.8$\times10^7$ \\
6 - 10  &  0.47              & 3.3         &  9.4$\times10^7$ & 9.4$\times10^7$ \\
10 - 12 &  3.5               & 3.8         &  1.9$\times10^6$ & 1.0$\times10^8$ \\ \hline 
\end{tabular}
\end{center}
\end{table}

The VLBI constraints show that the gyrosynchrotron compact source is
contained within a volume approximately 15~R$_{\odot}$ in radius.  The
strength of the magnetic field implicated in the maser strongly
suggests an origin for this emission close to the stellar surface, in
the region with strongest magnetic field strength. The value of
1.5-3kG is consistent with values found in sunspots. It is also
broadly consistent with values found using other techniques for
T~Tauri stars.  For example, Guenther et al. (1999) used a technique
based on Zeeman broadening of infrared Fe lines and determined
magnetic field strengths of 2.35~kG in the case of T~Tauri~N, and
1.1~kG for LkCa~15, a weak-lined T~Tauri star. Marginal detections of
kilogauss strength fields were made for two other stars, one classical
(accreting) and one weak-lined (non-accreting). It should be noted
that these fields represent the average field over the entire stellar
surface, so that local field strengths could be much higher.  Basri
et al. (1992) also used a similar technique to make one
detection and establish one upper limit on two weak-lined T~Tauri
stars.  Johns-Krull et al. (1999a) used spectropolarimetry of the
accreting T Tauri star BP Tau to detect a field of 2.4kG associated
with the He I $\lambda5876$ line, which is believed to form in a high temperature
accretion shock. A null result from photospheric lines indicates that
this strong field is not globally present and points to magnetically
funnelled accretion in this source.

Our observations point to strong magnetic fields and reconnection in
the immediate environment ($<15$R$_{\odot}$) of T~Tau~Sb. It is
tempting to speculate that this magnetic region is the actual field
linking the star with a disk, which would then be responsible also for
funnelling accreting material onto the star, and that the reconnection
is driven by field lines being twisted by the differential rotation of
star and inner disk. Could this be the case?  The persistent polarized
emission suggests that the emission arises in regions with magnetic
fields that are large scale and persistent on timescales of
$\sim$12~hours. We can impose a {\em minimum} scale size on the
emitting region, if we require that the emission before UT=6 hours is
gyrosynchrotron emission from non-thermal electrons, and impose a
limit on the maximum allowable brightness temperature.  The
gyrosynchrotron brightness temperature is limited by the mean energy of
the radiating particles.  For a maximum allowed T$_b$ of $10^{9}$~K
the minimum source radius would be approximately 3~R$_{\odot}$. If we allow
T$_{b}$ to be as high as $10^{10}$~K, requiring electrons in the MeV
range, the minimum source radius would be about 1~R$_{\odot}$.  For
the emission between UT=7 hours and UT=10 hours, the minimum source
size is 4~R$_{\odot}$ for T$_b$=10$^{9}$~K or 1.5~R$_{\odot}$ for
T$_b$=10$^{10}$~K.  The expected size of a T~Tauri magnetosphere can
be roughly estimated from the assumed field strength and accretion rate,
e.g. from Clarke et al. (1995)
\begin{equation}
R_m=\left(\frac{2B_*^2 R_*^6}{(G M_*)^{1/2} \dot{M}}\right)^{2/7} .
\end{equation}
Taking $B_*$=1.5~kG, as estimated above, $R_*=1.5 R_{\odot}$,
$M_*=0.25 M_{\odot}$, typical for a pre-main sequence M-dwarf, and
$\dot{M} = 10^{-7}~M_{\odot}$~yr$^{-1}$, which is at the high end of
the range of typical classical T Tauri accretion rates, we would
expect $R_m \sim 15~R_{\odot}$. Thus the magnetic field of T~Tau~Sb
has a scale size consistent with the expected size of a classical T
Tauri magnetosphere.

\section{Conclusion}

Our 8.4GHz observation of T~Tau suggests three types of emission:
Thermal emission from wind sources that are extended, a slowly varying
gyrosynchrotron source and a highly-polarized maser source.

In the high resolution VLBI data, only one pointlike source is
observed.  We identify this as the southern component T~Tau~Sb.
Intercontinental VLA-Effelsberg baselines in the early part of the
experiment allow us to constrain the size of this source to be less
than 14.5~R$_{\odot}$ in radius.  Circular polarization indicates the
presence of magnetic fields in the source region. Short timescale
variability is observed, in the form of two distinct step-like flux
increases, each followed by steady increased flux. During the first of
these, the observed polarization changed from left- to
right-handed. The second seems to have added only right-hand polarized
flux. We argue that this strongly indicates coherent emission, most
probably an electron-cyclotron maser. Based on this interpretation the
magnetic field strength in the maser emitting region can be estimated
to be in the kilogauss range.

By arguing that the quiescent emission in the
first 6 hours represents gyrosynchrotron emission from non-thermal
electrons and requiring the brightness temperature for this emission
to lie below sensible limits, we argue that the emitting magnetic
region must be at least 1~R$_{\odot}$ in size, and probably
larger. The gyrosynchrotron emitting regions have sizes of
the same order as those 
expected for an accreting magnetosphere.

At low resolution the system is resolved into the two components, N
and S.  T~Tau~N is consistent with a point source in our VLA maps.
T~Tau~S shows evidence of resolved flux, indicating that there is
either diffuse emission, or additional components in the system, or
both.  The non-detection of T~Tau~N in the high resolution data
indicates that this is a wind source. The non-detection of T~Tau~Sa
may indicate that this source is below our detection limit, although
it is possible that it is resolved out by our shortest VLBI
baseline.

\begin{acknowledgements}

We thank Stephen White and Ken Johnston for helpful discussions
concerning miscellaneous radio data, and the referee, T. Ray, for his
comments which allowed us to improve the manuscript. We also thank the
data analysts and staff at NRAO for their support throughout this
project.

\end{acknowledgements}


\begin{thebibliography}{}



  \bibitem[1992]{basri} Basri G., Marcy G.W. \& Valenti J.A., 
1992, ApJ  390,  622

  \bibitem[2002]{benz} Benz A.O.,
{\it Plasma Astrophysics},
second edition, Astrophysics and Space Science Library, 
Vol. 279, Kluwer Academic Publishers, Dordrecht, 2002. 

  \bibitem[1994]{benzguedel} Benz A.O. \& G\"udel M., 1994, 
A\&A 285, 621

  \bibitem[1995]{clarke} Clarke C.J., Armitage P.J, Smith K.W. \& Pringle J.E.,
1995, MNRAS 273, 639

  \bibitem[2002]{duchene} Duch\^ene G., Ghez A.M. \& McCabe C.,  
2002, ApJ 568, 771

  \bibitem[1982]{dyck} Dyck H.M., Simon T. \& Zuckerman B.,  
1982, ApJ 255, 103

  \bibitem[1978]{elias} Elias J.H.,  
1978, ApJ 224, 857

  \bibitem[1991]{ghez} Ghez A. M., Neugebauer G., Gorham P. W., et al.,
1991, AJ 102, 2066.

  \bibitem[1999]{guenther} Guenther E.W., Lehmann H., Emerson, J.P. \&
Staude J., 1999, A\&A 341, 768

  \bibitem[1996]{Hayashi} Hayashi M.R., Shibata K. \& Matsumoto R., 1996, ApJ 468, L37

  \bibitem[1999]{johnskrull2} Johns-Krull C.M., Valenti J.A., Hatzes A.P.  \&
  Kanaan A., 1999a, ApJ 510, L41.

  \bibitem[1999]{johnskrull1} Johns-Krull C.M., Valenti J.A. \& Koresko C., 
1999b, ApJ 516, 900

  \bibitem[2003]{johnston} Johnston K.J., Gaume R.A., Fey A.L., de Vegt C. \& Claussen M.J.,
2003, ApJ 125, 858

  \bibitem[2002]{kastner} Kastner J.H., Huenemoerder D.P., Schulz N.S.,  Canizares C.R.,
 \& Weintraub D.A., 2002, ApJ 567, 434

  \bibitem[2000]{kohler} K\"ohler R., Kasper M. \& Herbst T.,  
2000, In {\em Birth and evolution of Binary stars} poster proceedings 
of IAU Symp. 200, Reipurth \& Zinnecker (eds) p63.

  \bibitem[2000]{koresko} Koresko C.D.,  
2000, ApJ 531, L147

  \bibitem[1994]{vanlangevelde} van Langevelde H.J, van Dishoeck E.F., van der Werf P.P. \& Blake G.A.
1994, A\&A 287, L25

  \bibitem[1996]{momose} Momose M., Ohashi N., Kawabe R., Hayashi M. \& Nakano T.,  
1996, ApJ 470, 1001

  \bibitem[2000]{montmerle} Montmerle T., Grosso N., Tsuboi Y. \& Koyama K.,
2000, ApJ 532, 1097

  \bibitem[1990]{morris} Morris D.H., Mutel R.L. \& Su B.,
1990, ApJ 362, 299

  \bibitem[1993]{philips} Philips R.B., Lonsdale C.J. \& Feigelson E.D., 
1993, ApJ 403, L43

  \bibitem[1997]{ray} Ray, T.P., Muxlow T.W.B., Axon D.J., et al.,
1997, Nature 385, 415

  \bibitem[2000]{roddier} Roddier, F., Roddier C., Brandner W., Charrissoux D.,
V\'eran J.-P. \& Courbin F., 
2000, In IAU Symp. 200 {\it Birth and Evolution of Binary Stars}, poster proceedings,
Reipurth \& Zinnecker (eds), p60.

  \bibitem[1994]{skinner} Skinner, S.L. \& Brown A., 
1994, AJ 107, 1461

  \bibitem[1999]{white} White, S.M. 2000, In 
{\it Radio Interferometry: The Saga and the Science},
        ed. DG Finley, W Miller Goss,
        NRAO workshop no 27,
        pp. 86--111.
        Associated Universities, Inc.

  \bibitem[2001]{wood} Wood K., Smith D., Whitney B., et al.,
2001, ApJ 561, 299


\end{thebibliography}
\end{document}